# Subunit rotation in single FRET-labeled $F_1$-ATPase hold in solution by an anti-Brownian electrokinetic trap

Hendrik Sielaff[a], Thomas Heitkamp[a], Andrea Zappe[b], Nawid Zarrabi[b], Michael Börsch[a,b,*]

[a] Single-Molecule Microscopy Group, Jena University Hospital, Friedrich Schiller University Jena, Nonnenplan 2 - 4, 07743 Jena, Germany
[b] 3rd Institute of Physics, University of Stuttgart, Pfaffenwaldring 57, 70550 Stuttgart, Germany

**ABSTRACT**

$F_oF_1$-ATP synthase catalyzes the synthesis of adenosine triphosphate (ATP). The $F_1$ portion can be stripped from the membrane-embedded $F_o$ portion of the enzyme. $F_1$ acts as an ATP hydrolyzing enzyme, and ATP hydrolysis is associated with stepwise rotation of the γ and ε subunits of $F_1$. This rotary motion was studied in great detail for the last 15 years using single $F_1$ parts attached to surfaces. Subunit rotation of γ was monitored by videomicroscopy of bound fluorescent actin filaments, nanobeads or nanorods, or single fluorophores. Alternatively, we applied single-molecule Förster resonance energy transfer (FRET) to monitor subunit rotation in the holoenzyme $F_oF_1$-ATP synthase which was reconstituted in liposomes. Now we aim to extend the observation times of single FRET-labeled $F_1$ in solution using a modified version of the anti-Brownian electrokinetic trap (ABELtrap) invented by A. E. Cohen and W. E. Moerner. We used Monte Carlo simulations to reveal that stepwise FRET efficiency changes can be analyzed by Hidden Markov Models even at the limit of a low signal-to-background ratio that was expected due to high background count rates caused by the microfluidics of the ABELtrap.

## 1 INTRODUCTION

$F_oF_1$-ATP synthase is an ubiquitous membrane protein that utilizes the electrochemical potential of protons over the membrane, that is the proton motive force (PMF), to synthesize adenosine triphosphate (ATP) from adenosine diphosphate (ADP) and inorganic phosphate ($P_i$). The bacterial enzyme can also work in reverse. Depending on the physiological conditions $F_oF_1$-ATP synthase hydrolyzes ATP in order to pump protons across the membrane[1]. Here, we focus on the soluble $F_1$ portion where ATP hydrolysis takes place. In its simplest bacterial form $F_1$ consists of five different subunits, namely $\alpha_3\beta_3\gamma\delta\epsilon$. The structure of *Escherichia coli* $F_1$ was recently resolved by a crystal structure at a resolution of 3.15 Å (Fig. 1A)[2, 3]. The main body of $F_1$ consists of a pseudohexagonal structure formed by three pairs of subunits α and β, $\alpha_3\beta_3$. Each subunit β provides a nucleotide binding site that can bind and catalyze ATP synthesis or hydrolysis, while the corresponding nucleotide binding sites in each α subunit are catalytically inactive. Subunits $\alpha_3\beta_3$ together with subunit δ at the top of $F_1$ provide the rigid stator that stabilizes the complex[4]. In contrast, subunits γ and ε form the central stalk which can rotate[5]. The globular portion of subunit γ together with subunit ε connect to the membrane-embedded ring of *c*-subunits of the $F_o$ domain. On the other side the C- and N terminal α-helices of subunit γ intrude the $\alpha_3\beta_3$ pseudohexagon. Thereby, the holoenzyme can transfer the energy of the proton motive force generated in $F_o$ *via* rotational movements of c-ε-γ to the nucleotide binding sites in $\alpha_3\beta_3$, where it is transformed into chemical energy of ATP.

In order to maintain a high kinetic efficiency for energy conversion subunit γ must be elastically flexible[6, 7]. It was shown that γ is a least ten times more flexible than the stator. The domain with highest elasticity is located at the interface to the *c*-subunits where the two rotor portions of $F_o$ and $F_1$ are connected[8, 9]. Subunit γ transfers energy by its curved N-terminal α-helix to the DELSEED-sequence in subunit β, a helix-turn-helix motive, that interacts with the nucleotide binding site. Depending on the orientation of subunit γ the nucleotide binding sites are sequentially opened and closed, synchronizing binding and hydrolysis of ATP, and releasing the products ADP and $P_i$.

\* Email:   michael.boersch@med.uni-jena.de   or   m.boersch@physik.uni-stuttgart.de   ;   http://www.m-boersch.org

The exact sequence of events was revealed by a series of single molecule fluorescence microscopy experiments (see review[10]). The $\alpha_3\beta_3\gamma$ part of $F_1$ was bound *via* Histidine-tags in each $\beta$ subunit to the glass surface, while a fluorescent probe (actin filament, polystyrene bead, or magnetic bead) was attached on the opposite side of subunit $\gamma$ to monitor its rotational movement during ATP hydrolysis. Subunit $\gamma$ showed a 120° stepped rotation at high ATP concentrations corresponding to the three-fold symmetry of $\alpha_3\beta_3$. Furthermore, each 120° step consists of a 80° and a 40° substep[11, 12]. In a sophisticated experimentally setup with additionally fluorescently labeled ATP the ATP binding reaction was correlated with the 80° substep. Further experiments showed that ATP is hydrolyzed in a pause between the two substeps, and that the 40° substep is related to the exergonic release of $P_i$ from the nucleotide binding site. Finally ADP is released before the start of the next reaction cycle.

In order to prevent ATP hydrolysis *in vivo* the bacterial $F_oF_1$-ATP synthase is thought to be regulated by subunit $\varepsilon$, a 15 kDa subunit which is part of the $F_1$ rotor[13]. It consists of an N-terminal $\beta$-sandwich domain that binds to the globular domain of subunit $\gamma$ and to the *c*-ring, and a C-terminal domain with two $\alpha$-helices[14, 15] in an 'extended'-configuration (Fig. 1A) or a distinct 'up'-configuration, respectively (Fig. 1B). It is considered to be an intrinsic inhibitor of the $F_oF_1$-ATP synthase. However, in the active enzyme the C-terminal domain is thought to form a hairpin-folded state[16] with the C-terminal helices in a 'down'-configuration (Fig. 1C). The two $\alpha$-helices can extend parallel to subunit $\gamma$ into the cavity of $\alpha_3\beta_3$, as shown by the crystal structures[2, 17] (Fig. 1A and B), and thereby inhibit ATP hydrolysis activity by stalling the rotor at a fixed angle. A higher activation energy is needed to reactivate the enzyme from this $\varepsilon$-inhibited state than from another inhibited state, the so-called MgADP-inhibited state[18]. Recently, an *E. coli* enzyme with a deleted C-terminal domain of subunit $\varepsilon$ showed not only a higher ATP hydrolysis (ATPase) activity compared to the wild type, but also a higher ATP synthesis activity, that was not observed before[19]. The authors suggest that the C-terminal domain in its extended form suppresses multiple elementary steps of the ATP synthesis or hydrolysis reactions that are executed in the three $\beta$ subunits. In particular the C-terminal domain blocks the rotation of subunit $\gamma$, and, accordingly, all reactions that are accompanied by rotation, i.e. all rates of product release. On the other hand, the rates of covalent bond formation or cleavage of ATP are unaffected by subunit $\varepsilon$ inhibition. In addition, biochemical experiments performed by S. D. Dunn and coworkers[20] suggest that $\varepsilon$ interacts also directly with the $\beta$ subunits. Therefore it is believed that the physiological role of subunit $\varepsilon$ is the inhibition ATP hydrolysis of the $F_oF_1$-ATP synthase when the proton motive force and the ATP concentration in the cell are low to prevent waste of ATP.

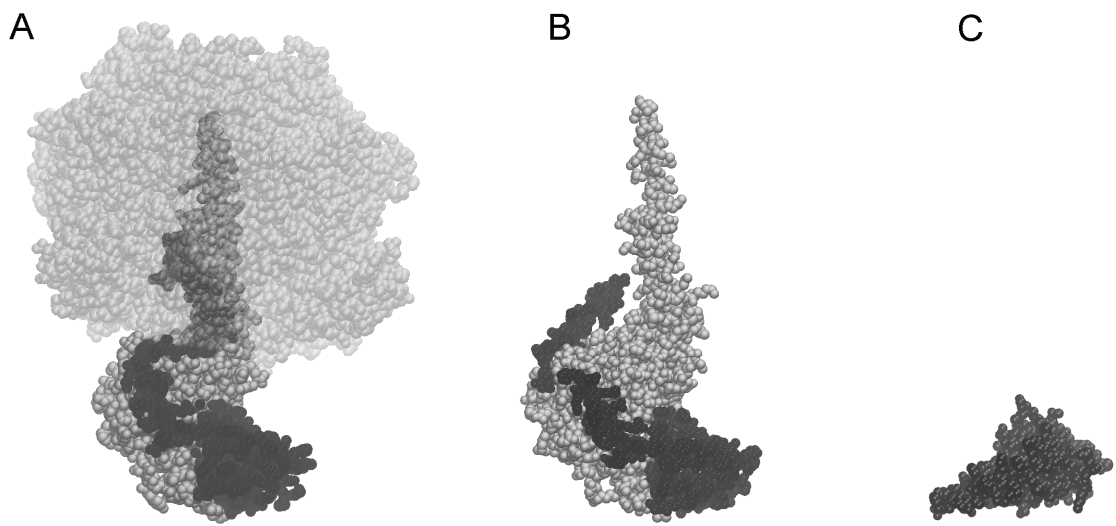

**Figure 1.** (**A**) Structure of *E. coli* $F_1$ with subunits $\alpha_3\beta_3$ in 'transparent' grey, $\gamma$ in grey and $\varepsilon$ in black[2]. (**B**) Partial structure of the $\gamma$-$\varepsilon$ subunits complex from *E. coli* with the C-terminal helices of $\varepsilon$ in the 'up'-conformation[21]. (**C**) Partial structure of the $\varepsilon$-subunit from *E. coli* $F_1$ with the C-terminal helices in the 'down'-configuration[15].

Our group investigates conformational changes and subunit rotation in single $F_oF_1$-ATP synthase using the Förster resonance energy transfer (FRET) approach since 1997[22-51]. We have developed a specific FRET labeling scheme with both fluorophores attached to the $F_1$ portion to monitor rotation of subunit γ [24]. However, a major limitation of our confocal single-molecule FRET approach using freely diffusing, liposome-reconstituted $F_oF_1$-ATP synthase is the short observation time, from of a few milliseconds for $F_1$ to some hundred milliseconds for reconstituted $F_oF_1$. Recently we have built an Anti-Brownian Electrokinetic Trap (ABELtrap, invented by A. E. Cohen and W E Moerner) that can hold single small particles like 20-nm fluorescent beads for more than 8 seconds. These fast ABELtraps capture liposomes, DNA, proteins or even single fluorophores in solution[52-54] with very fast feedback times in the microsecond time range[55-59]. Here, we describe the current status of our $F_1$-ATPse preparations for future single-molecule FRET measurements. We reveal the recovery of FRET levels by Hidden Markov Models (HMMs) at the lower limit of signal-to-background ratios, as expected in single-molecule FRET data using low laser excitation power for maximum observation times of the enzyme in the ABELtrap.

## 2 EXPERIMENTAL PROCEDURES

### 2.1 Preparation and characterization of $F_1$-ATPase

The $F_1$ portion ($F_1$-ATPase, $F_1$) of $F_oF_1$-ATP synthase is prepared from the plasma membranes of *Escherichia coli*. Details of our actual preparation procedures are given below.

*Construction of plasmid pMB3*

Plasmid pMD2 (M. G. Düser, published in [45]) is a derivative of pACWU1.2 [60] and carries a V78C mutation in the *c*-subunit and an N-terminal 6×Histidine-tag with the extension MRGS-HHHHHH-G- in the β-subunit of $F_1$-ATPase. Plasmid pMB6 is a derivative of plasmid pRA100 [61] and carries a 56C point mutation in the ε-subunit[27]. Plasmid pMD2 was digested with the restriction enzymes PmeI and SacI to yield a 548 bp long fragment carrying the 6×Histidine-tag at the N-terminus of subunit β. Plasmid pMB6 was digested with the same restriction enzymes, and a 12154 bp long fragment was isolated were the N-terminal sequence of the β-subunit was deleted. Both fragments were ligated resulting in the plasmid pMB3.

*Bacterial strains and growth conditions*

For the expression of the *E. coli* atp genes, strain RA1 *(F⁻ thi rpsL gal Δ(cyoABCDE)456::KAN Δ(uncB-uncC) ilv:Tn10)*[62] was used, which lacks a functional $F_oF_1$-ATP Synthase. This strain was transformed with the plasmid pMB3. Cells were grown in a modified complex medium (0.5 g/l yeast extract, 1 g/l tryptone, 17 mM NaCl, 10 mM glucose, 107 mM $KH_2PO_4$, 71 mM KOH, 15 mM $(NH_4)_2SO_4$ and 4 μM uracil, 50 μM $H_3BO_3$, 1 μM $CoCl_2$, 1 μM $MnCl_2$, 2 μM $ZnCl_2$, 10 μM $CaCl_2$, 3 μM $FeCl_2$, 0.5 mM $MgSO_4$, 0.5 mM arginine, 0.5 mM isoleucine, 0.7 mM valine, 4 μM thiamine, 0.4 μM 2,3-dihydroxybenzoic acid) at 37° C in a 10 L FerMac 320 fermenter (Electrolab, UK), harvested in the late logarithmic phase, and pelleted in a Sorvall Evolution RC centrifuge (Thermo Fisher Scientific, USA) at 10,000 × g, 4 °C, for 5 min.

*Purification of $F_1$-ATPase using Histidine-tags*

$F_oF_1$-ATP synthase-containing membranes were purified and $F_1$-ATPases were stripped off according to[63] with some minor modifications. The cell lysis buffer contained 10 % (v/v) glycerol instead of the sucrose originally used. Furthermore, cells were lysed by two passages through a PandaPlus 2000 cell homogenizer (GEA Niro Soavi, Italy) at 1000 bar. Finally, instead of a PEG6000 precipitation, the soluble $F_1$ portion was precipitated by the addition of 65 % (v/v) saturated $(NH_4)_2SO_4$. The precipitated and pelleted $F_1$ was resuspended in buffer A (20 mM Tris-HCl pH 8, 300 mM NaCl, 1 mM $MgCl_2$, 5% (v/v) glycerol, 10 mM imidazole, 0.25 mM PMSF), loaded on a HisTrap FF column (GE-Healthcare, USA) equilibrated with buffer A, and connected to an Äkta PrimePlus FPLC system (GE-Healthcare, USA). After washing the column with five column volumes of buffer A, $F_1$ was eluted by an imidazole gradient over 20 column volumes up to 0.5 M imidazole in buffer A. $F_1$-containing peak fractions were pooled and precipitated with 65 % (v/v) saturated $(NH_4)_2SO_4$. Pelleted $F_1$ was then resuspended in 0.5 ml buffer B (buffer A with 150 mM NaCl), and loaded on a Superdex 200 10/300 GL size exclusion chromatography column (GE-Healthcare, USA) equilibrated with buffer B and eluted using an Äkta PrimePlus FPLC system (GE-Healthcare, USA) with a flow rate of 0.5 ml/min.

The major peak fractions containing $F_1$ were pooled, concentrated using an Amicon Ultra-15 Ultracell 30K (Merck Millipore, USA), shock-frozen in liquid nitrogen in 500 µl cryo straws, and stored at -80 °C.

*Measurement of ATP hydrolysis activity*

ATPase activity was measured on the basis of established protocols[64]. Briefly, 2 µl of purified protein containing 1-2 µg of the enzyme was added to buffer C (0.5 ml 50 mM Tris-Acetic acid pH 8.5, 10 mM ATP, 4 mM Mg-acetate), and incubated at 30 °C for 5 min. The reaction was stopped by adding 0.5 ml of 10 % (w/v) SDS and the free phosphate was determined by measuring absorbance at 700 nm after adding 0.5 ml ferrous ammonium sulfate-$H_2SO_4$ ammonium molybdate reagent[65]. Specific LDAO activation of ATPase activity was determined by adding 0.5 % (v/v) LDAO to the reaction mixture.

*Other methods*

Protein concentrations were determined using published methods[66]. SDS-polyacrylamide gels were made as described[67] with acrylamide concentrations between 12 - 14 %. Silver staining was performed according to[68].

**2.2 Simulation of FRET data mimicking subunit rotation in $F_1$-ATPase**

We aim at prolonged observation times of single FRET-labeled $F_1$-ATPase in solution using an ABELtrap. To evaluate the minimum signal-to-background ratio required for subsequent FRET data analysis by Hidden Markov Models (HMMs), we simulated the stepwise rotation of subunit γ of $F_1$ using a Monte Carlo simulation. Therefore, a single FRET-labeled particle was placed into a virtual box. A threedimensional ellipsoid was centered within this box representing the threedimensional detection volume of a confocal experiment. Previously the FRET-labeled particle could diffuse freely in an out of the ellipsoid[31]. Here, we changed the simulations so that the FRET-labeled particle was confined inside the ellipsoid as soon as it diffused into the boundaries of the ellipsoid. Depending on the given photophysical rates for photobleaching of the FRET donor and FRET acceptor dyes, the particle was forced to stay inside the ellipsoid until the fluorescence signal disappeared irreversibly. Subsequently the next freely-diffusing FRET-labeled particle was generated in the box but outside of the ellipsoid, and the ABELtrap simulation proceeded. Low photon count rates with shot-noise-limited intensity fluctuations plus a high background count rate on both detection channels were used yielding nearly constant count rates for the sum of both FRET donor and acceptor fluorescence photons.

**2.3 Hidden Markov Model-based FRET level analysis**

The simulated FRET time trajectories of a single particle in the ABELtrap were analyzed using previously described Hidden Markov Models (HMM) with the given number of 4 states. Three different FRET levels and one 'donor only' state (after FRET acceptor photobleaching) were expected. In contrast to our previous HMM approach to find FRET states in freely-diffusing proteoliposomes[31, 34] we used Gaussian distributions for each FRET state with variable widths. After the first round of HMM learning the 3 FRET levels, the 'donor only' state and the associated dwell times, the resulting FRET levels and corresponding dwell times were used to assign these FRET levels (and the 'donor only' state) to the FRET time trajectories. The dwell time histograms of the assigned FRET levels were fitted with monoexponential decay functions to unravel the deviations between the learned HMM values and the assigned FRET levels.

# 3  RESULTS

The purification procedures of a fully active $F_1$-ATPase from *E. coli* were established and described briefly. In contrast to our recently published $F_oF_1$-ATP purification (T. Heitkamp et al., Proc. SPIE 8588 (2013)) which included the use of a new 10 L fermenter system for cell growth and a different type of cooled cell homogenizer to prepare the plasma membranes, we purified $F_1$-ATPase with Histidine-tags using Ni-NTA column chromatography on an Äkta Prime Plus FPLC system. Cell membranes were collected after lysis of the cells by disruption at 1000 bar. The $F_1$ portion of the $F_oF_1$-ATP synthase was purified from cell membranes in two steps. First, samples were loaded on a HisTrap column binding the $F_1$-ATPase *via* its Histidine-tags in each β-subunit to the matrix of the column before elution with imidazole. In the second step, protein-containing fractions were applied to a size exclusion chromatography on a Superdex 200 column that served as a final polishing step to remove all remaining protein impurities. This procedure yielded $F_1$-ATPase of high purity as shown by SDS-PAGE in Fig. 2. The final protein sample contained only the $F_1$ subunits α, β, γ, δ, and ε, but not the $F_o$ subunits *a*, *b*, and *c*. Most of the $F_o$ subunits and other membrane proteins were stripped from the $F_1$ portion on the HisTrap column, where $F_1$ was bound to the column matrix, but not $F_o$. Now, single-molecule FRET experiments on the rotation of subunits γ and ε as well as the conformational changes of the C-terminus of ε are possible.

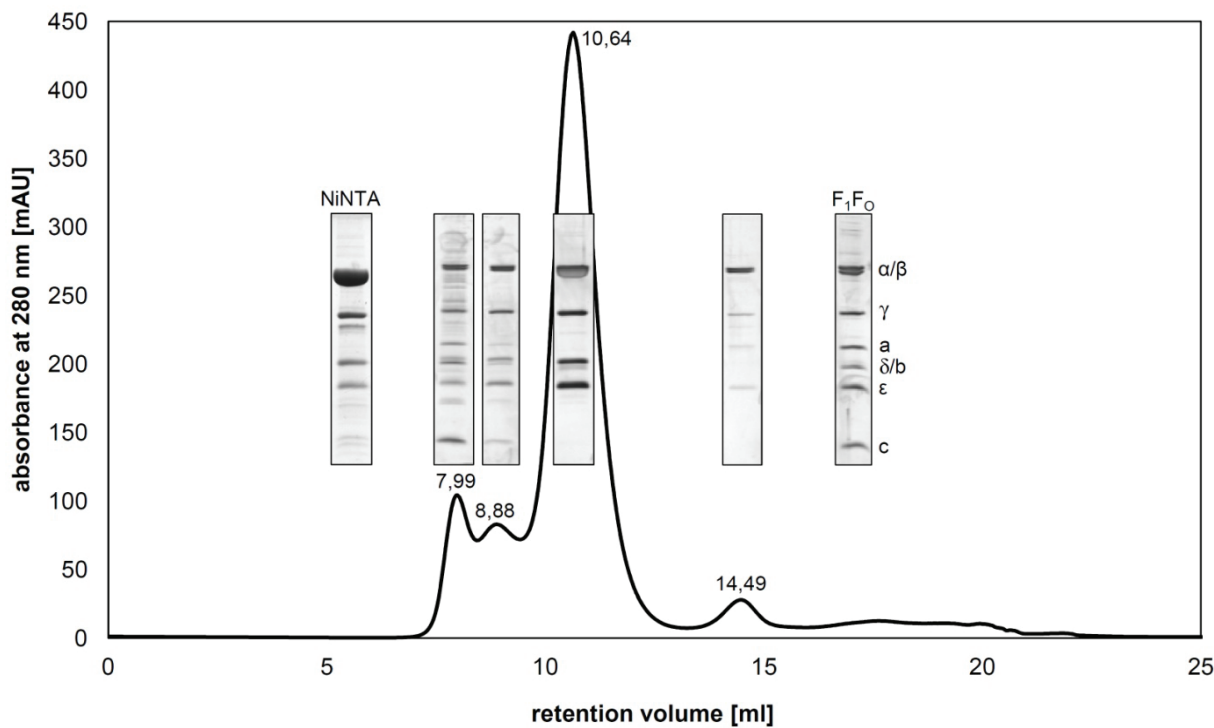

**Figure 2.** Purification profile of $F_1$-ATPase from *E. coli* (i.e. size exclusion chromatography after HisTrap column purification of $F_1$-containing fractions). Fractions with the highest protein concentrations were pooled, concentrated, and afterwards injected on a Superdex 200 column. The retention volumes of the peaks are indicated. Furthermore, the corresponding silver-stained SDS-PAGE of every peak fraction are shown. On the left, the SDS-PAGE of the injected sample is shown. The first protein peak (7.99 ml) corresponds to the void volume of the column and contained most of the impurities. The last protein peak on the right (14.49 ml) contained mainly the α and β subunits alone. The major peak contained $F_1$ with a high purity and without $F_o$ subunits *a*, *b*, and *c*. The SDS-PAGE (see lane on the far right) shows a $F_oF_1$ standard for a better identification of the $F_1$ subunits α, β, γ, ε and δ.

ATP hydrolysis activities (ATP turnover) of the purified $F_1$-ATPase were measured to reveal functionality. The activity was measured at 30° C with an assay that determines the concentration of the released phosphate photometrically. We found an activity for the wild type $F_1$-ATPase of about 260 s$^{-1}$, which is in the same range as published activities[69].

The observation time of the freely diffusing $F_1$-ATPase is approximately in the range of 1 to 3 ms using a confocal detection volume with a size of a few femtoliters. An ABELtrap is required to obtain FRET time trajectories of single $F_1$ captured in solution that last for several seconds. However, trapping FRET-labeled $F_1$ is not enough. Photobleaching of FRET donor or acceptor fluorophores depends on the number of excitation cycles. We aim to extend these observation time limitations by applying a reduced laser excitation power. This will result in a low photon count rate of both fluorophores. Therefore, we had to verify that the low photon count rates in the FRET time trajectories can be analyzed quantitatively using Hidden Markov Models. A Monte Carlo simulation of a single trapped FRET-labeled particle (like $F_1$) in a box was used to explore FRET data analysis with HMMs.

A single particle was placed inside a box with 2.6 µm × 2.6 µm in x and y dimensions, and 13.2 µm in z dimension. The particle moved due to Brownian motion and according to its size of 10 nm. The confocal threedimensional Gaussian detection volume with 5.8 fl (0.65 µm for x- and y-radii, 3.3 µm for z radius) was centered inside the box. Once the particle hit these boundaries it was considered 'trapped' and emitted on average 10000 photons per second for the FRET donor fluorophore as well as for the FRET acceptor fluorophore. Three different FRET efficiencies were set, at proximity factors 0.3, 0.5, and 0.7, and with short dwell times of 10 ms per level. A preferred sequence of 0.3→0.7→0.5→0.3→ was used to mimic unidirectional rotary movement, or sequential conformational changes, respectively. Photobleaching of the FRET donor was simulated based on a mean emission of 20000 photons before bleaching, and of 15000 photons before bleaching for the FRET acceptor. On both detection channels a mean photon count rate of 10000 counts per second (10 kHz) was added, similar to a background observed in experimental ABELtrap data. Ten FRET trajectories with 60 s duration each were simulated resulting in total 252 photon bursts with 8466 states.

For the subsequent HMM analysis, we binned the FRET time trajectories with 1 ms per data point. Data visualization was achieved with the software 'burst_analyzer'[28], and background correction was used with 10 kHz for both FRET donor and acceptor channels. The HMM with 4 states (3 FRET levels plus one 'donor only' state for particles with a photobleached FRET acceptor dye) were given, but no preferred sequence of states was implemented. The starting values of the HMM were set to proximity factors 0.2 (for 'donor only' state), 0.4, 0.6 and 0.8, with variances of 0.02 for each FRET level, and dwell times of 20 ms.

The result of the HMM after learning was promising. FRET levels were learned at 0.1133 (with variance 0.0018, that is the 'donor only' state), 0.3145 with variance 0.0083, 0.5028 with variance 0.0125, and 0.6892 with variance 0.0082. All FRET levels were found in very good agreement with the simulated values. The associated dwell times were learned with 3741 ms for the 'donor only' state, and 9.31 ms, 9.18 ms or 9.73 ms for the three FRET levels 0.31, 0.50 or 0.69, respectively. Again, the identification of the FRET levels was in very good agreement with the simulation.

The final step of the HMM approach is the assignment of these learned FRET levels and dwell times to the simulated FRET time trajectories. The results are shown in Fig. 3. The two examples of photon bursts in Fig. 3A and B were terminated by FRET donor photobleaching. In Fig. 3A, the FRET acceptor bleached before the donor, and the remaining 'donor only' state with apparent FRET efficiency around 0.1 became visible at the end of the time trajectory. The assignment of FRET levels by the HMM was in good agreement with the simulation despite the low count rates and the additional noise contribution after background corrections of the data. Manual assignment of FRET levels appeared to be difficult.

The FRET transition density plot[70] showed three pronounced transitions (Fig. 3C). The transitions from FRET level 1 at 0.7 to the following FRET level 2 at 0.5, and from 0.5 to 0.3, and from 0.3 to 0.7 clearly indicated the existence of a preferred sequence. This was the same sequence as the simulated one. The alternative transitions, that is from 0.3 to 0.5 to 0.7 to 0.3 and so on, were not assigned very often.

The dwell time distributions for the three assigned FRET levels were fitted with monoexponential decay functions from the maximum of the distribution yielding dwell times of 11.0 ms for FRET level 0.3 and 11.8 ms for FRET level 0.7 (data not shown). For FRET level 0.5 the fit to the dwell time distribution in Fig. 3D resulted in (15.4 ± 0.5) ms, which is slightly longer than the given dwell time of 10 ms for these states. Nevertheless, the dwell times were learned correctly by the HMM, and the subsequent assignment procedure prolonged the dwells due to missing some short dwell of a few milliseconds.

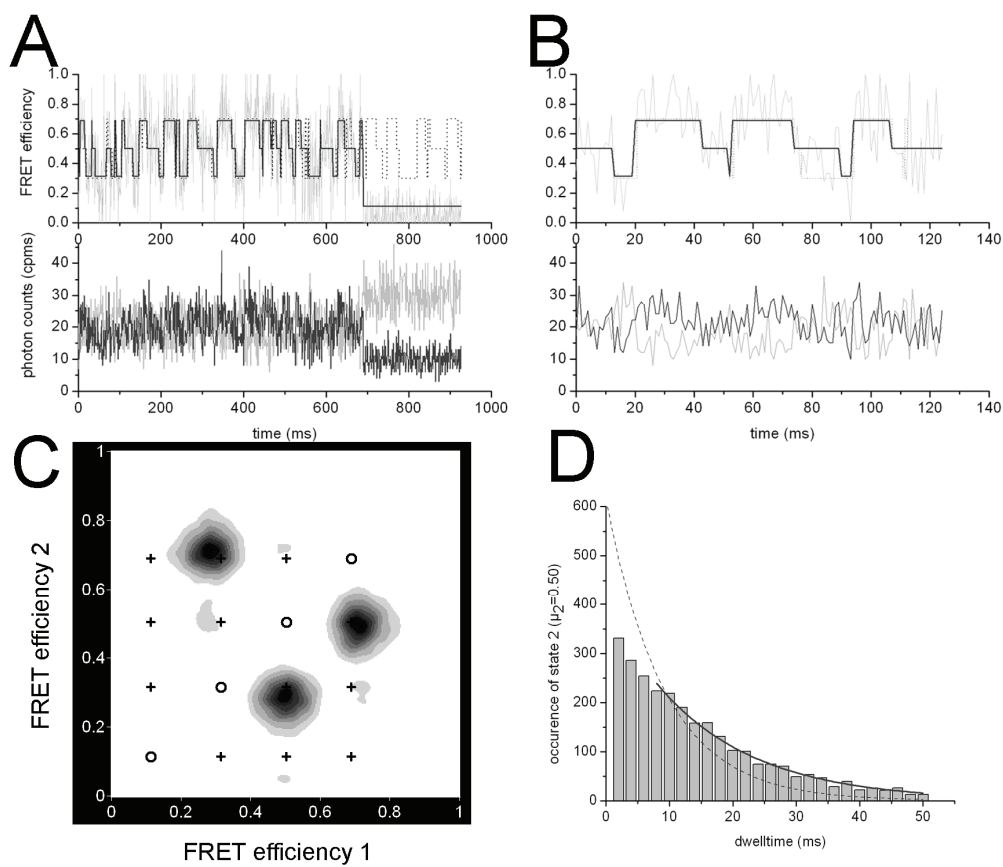

**Figure 3.** Simulated FRET data and FRET level recovery by a 4-state Hidden Markov Model. (**A**) Simulated FRET trajectory with FRET donor (grey line) and acceptor (black line) counts rates in counts per ms (cpms) in the lower panel, and the FRET efficiency trace in the upper panel. Recovered FRET levels are shown by the solid black line, simulated FRET levels by the dotted line. FRET acceptor photobleaching occurred after about 700 ms. (**B**) Simulated FRET trajectory with FRET donor and acceptor counts rates (lower panel) and FRET efficiency trace (upper panel). Assigned FRET levels are shown as a black line, simulated FRET levels as a dotted line. FRET donor photobleaching occurred after about 130 ms and terminated the photon burst. (**C**) FRET transition density plot from assigned FRET levels with three transitions clearly indicating the FRET level sequence. Crosses represent the expected transitions according to the simulation, open circles indicate apparent transitions with no change in FRET efficiency. (**D**) Dwell time distribution for the assigned FRET level $\mu_2$=0.50 in all 252 simulated bursts. The solid line is a monoexponential fit to the distribution, the dotted line is the expected dwell time distribution from the simulation.

## 4 DISCUSSION

Conformational changes of soluble proteins like $F_1$-ATPase can be monitored using two specifically attached fluorophores for an internal distance ruler based on FRET. Because surface attachment might perturb the conformational dynamics, single-molecule FRET data are measured in solution. However, the observation time of a 10-nm sized enzyme in a confocal detection volume is less than 5 ms due to Brownian motion. Therefore, time trajectories with several FRET transitions cannot be recorded in general.

We are interested in long time trajectories of FRET changes in $F_1$-ATPase for direct comparison with the holoenzyme, that is a liposome-reconstituted $F_oF_1$-ATP synthase, as well as with other ATP-driven membrane transporters[71-79]. As the first step we established a purification protocol for $F_1$-ATPase from *E. coli*. Histidine-tags added to the N-terminus of the β-subunits and the corresponding Ni-NTA-based chromatography were used for purification. The purified $F_1$ contained all five subunits and showed good ATP hydrolysis activity. Because we had shown previously how to label $F_1$ specifically with two fluorophores for FRET-based rotation measurements[24], we proceeded with improving single-molecule FRET analysis.

To identify the lower limit of the signal-to-background ratio for HMM-based FRET levels, we simulated FRET time trajectories with low photon count rates in the presence of a high background. These conditions were expected for capturing a single $F_1$-ATPase in solution by an ABELtrap[80]. The ABELtrap is a microfluidic device keeping the enzyme within the confocal laser focus until photobleaching of the fluorophores. Diffusion of the enzyme is limited to the x- and y-directions within a 1-μm shallow trapping region that is located between the cover glass and the PDMS chamber. Electrokinetic forces comprise electrophoretic and electroosmotic forces to counteract the actual movement of the protein in solution. Potentials are applied to four platinum electrodes to push back the protein to the center of the laser focus. We started with an ABELtrap approach using an EMCCD camera for particle localization[81]. Recently we have built a faster ABELtrap using a confocal laser pattern which is controlled by a field-programmable gate array (FPGA). This ABELtrap could hold 20-nm fluorescent polystyrene beads in solution for more than 8 seconds, that is with 1000-fold prolongation of the observation time (N. Zarrabi et al, Proc. SPIE 8587 (2013), in press).

The Monte Carlo simulations for fast FRET level changes with mean dwell times of 10 ms per FRET level provided 252 photon bursts with 8466 FRET levels for HMM analysis. The high background on both channels added more noise on the photon count trajectories after it was subtracted. The given 4-state HMM for the three FRET levels plus the additional 'donor only' state recovered these FRET levels precisely, and also the learned dwell times matched the simulation. The subsequent assignment of the learned FRET levels and dwells yielded the correct FRET transitions und also the predefined FRET level sequence of the simulation. Minor deviations were found for the assigned dwell times. Obviously the assignment of very short FRET levels of a few milliseconds is difficult at low signal-to-background, and, accordingly, these FRET levels were overlooked.

Two different experimental requirements for future FRET analysis of freely diffusing $F_1$-ATPase have been achieved. The enzyme was purified with all subunits and was fully functional for ATP hydrolysis, and a low signal-to-background FRET data set could be analyzed successfully by Hidden Markov Models. Therefore, we expect that measuring conformational dynamics like subunit rotation of γ/ε or regulatory conformational changes of the C-terminus of ε will be possible soon. Long FRET time trajectories obtained in an ABELtrap will provide the required statistical basis for a quantitative description of the rates for these conformational dynamics, measured with one active enzyme at a time.

**Acknowledgements**

This work was supported in part by DFG grants BO 1891/10-2 and BO 1891/15-1 to M.B.. Additional financial support by the Baden-Württemberg Stiftung (by contract research project P-LS-Meth/6 in the program "Methods for Life Sciences") is gratefully acknowledged. The authors want to thank Prof. Dr. T. Duncan (SUNY Upstate Medical University, Syracuse, NY, USA), Prof. Dr. S. D. Dunn (University of Western Ontario, London, Ontario, Canada) and Prof. Dr. P. Gräber (University of Freiburg, Germany) for their support for $F_1$ and $F_oF_1$ mutants and enzyme purifications.